\newif\ifmnras
\mnrasfalse
\ifmnras
	\documentclass[fleqn,usenatbib]{mnras}
\else
	\documentclass[apj,numberedappendix]{emulateapj}
\fi

\usepackage{graphics,epsf}
\usepackage{amsmath}                
\usepackage{amsfonts}               
\usepackage{amssymb}                
\usepackage{epsfig}                 
\usepackage{graphicx}
\usepackage{xcolor}

\definecolor{redak}{rgb}{0.9,0.15,0.05}

\def \cm{~\rm{cm}}
\def \s{~\rm{s}}
\def \km{~\rm{km}}

\def \K{~\rm{K}}
\def \g{~\rm{g}}

\def \erg{~\rm{erg}}

\def \yr{~\rm{yr}}

\ifmnras
	\def \aap{A\&A}
	
	\def \apj{ApJ}
	\def \apjl{ApJ}
	\def \apjs{ApJS}

	\def \mnras{MNRAS}

	 \def \na{New Astronomy}
\fi
\def \na{New Astronomy}
\ifmnras
\title[Jets in the wind acceleration zone of a giant star]
{A companion star launching jets in the wind acceleration zone of a giant star}

\author	[Hillel, Schreier, Soker]{
Shlomi Hillel$^{1}$, Ron Schreier$^{1}$, and Noam Soker$^{1,2}$\thanks{E-mail: shlomi.hillel@gmail.com; ronsr@physics.technion.ac.il; soker@physics.technion.ac.il}
\\
$^{1}$Department of Physics, Technion -- Israel
Institute of Technology, Haifa 32000 Israel\\
$^{2}$Guangdong Technion Israel Institute of Technology, Shantou 515069, Guangdong Province, China}

	\date{Accepted XXX. Received YYY; in original form ZZZ}

	\pubyear{2017}
\fi

\begin{document}
\label{firstpage}

\ifmnras
	\pagerange{\pageref{firstpage}--\pageref{lastpage}}
	\maketitle
\else
	\title{A companion star launching jets in the wind acceleration zone of a giant star}

	\author{Shlomi Hillel}
	\author{Ron Schreier}
	\author{Noam Soker}
	\affil{Department of Physics, Technion -- Israel
	Institute of Technology, Haifa 32000, Israel; shlomi.hillel@gmail.com; ronsr@technion.ac.il;
	 soker@physics.technion.ac.il}
\fi

\begin{abstract}
By conducting three-dimensional (3D) hydrodynamical simulations we find that jets that a main sequence companion launches as it orbits inside the wind acceleration zone of an asymptotic giant branch (AGB) star can efficiently remove mass from that zone. 
We assume that during the intensive wind phase a large fraction of the gas in the acceleration zone does not reach the escape velocity. Therefore, in the numerical simulations we blow the wind with a velocity just below the escape velocity. We assume that a main sequence companion accretes mass from the slow wind via an accretion disk, and launches two opposite jets perpendicular to the equatorial plane. This novel flow interaction shows that by launching jets a companion outside a giant star, but close enough to be in the acceleration zone of a slow intensive wind, can enhance the mass loss rate from the giant by ejecting some gas that would otherwise fall back onto the giant star. The jets are bent inside the wind acceleration zone and eject mass in a belt on the two sides of the equatorial plane. The jet-wind interaction contains instabilities that mix shocked jets' gas with the wind, leading to energy transfer from the jets to the wind. 
As well, our new simulations add to the rich variety of jet-induced outflow morphologies from evolved stars.
\end{abstract}

 \begin{keywords}
{ stars: AGB and post-AGB -- binaries: close -- stars: jets }
 \end{keywords}


\section{INTRODUCTION}
\label{sec:intro}

There is a rich variety of stellar binary systems that at one or more periods along their evolution one star accretes mass from the other star through an accretion disk and launches jets. Most relevant to the present study are systems where a main sequence companion accretes mass from a cool giant and launches jets  (e.g., \citealt{Wittetal2009, Thomasetal2013, Gorlovaetal2015, VanWinckel2017}). The jets themselves might in return influence the ambient gas that serves as the mass reservoir for the accretion disk. In particular, the jets might shape the outflow and enhance the mass loss rate from the binary system as to reduce the mass accretion rate and therefore the jets' power. By this the removal of mass by the jets sets a negative feedback cycle, the jet feedback mechanism (JFM; \citealt{Soker2016Rev} for a review). 

Alongside the negative component of the feedback cycle, the jets have a positive feedback component. The mass that flows onto the accreting compact object builds a high pressure zone near that object. This high pressure acts to reduce the accretion rate (e.g., \citealt{RickerTaam2012, MacLeodRamirezRuiz2015}). By removing angular momentum, high entropy gas, and energy from the close surroundings of the accreting compact object the jets reduce the pressure in that region and by that enable high mass accretion rate \citep{Shiberetal2016, Staffetal2016MN, Chamandyetal2018}.  
 
 In one type of binary systems where jets are thought to remove mass, a neutron star companion launches jets inside the envelope of a giant star (e.g., \citealt{ArmitageLivio2000, Chevalier2012, Papishetal2015, SokerGilkis2018, Gilkisetal2019}) or in its inflated unstable envelope (e.g., \citealt{DenieliSoker2019}). 
 White dwarfs (e.g., \citealt{Soker2004}) and in particular main sequence stars (e.g., \citealt{Soker2004, MorenoMendezetal2017, Sabachetal2017}) might also launch jets while spiraling-in inside the envelope of a giant star, and by that facilitate the removal of the common envelope (e.g., \citealt{LopezCamaraetal2019, Shiberetal2019}).  

In the present study we simulate the launching of jets outside the envelope of the giant star. We are interested in cases where the dense wind of the giant star, before the wind reaches its terminal velocity, overflows its Roche lobe, but  the giant envelope does not overflow its Roche lobe. If the companion is closer to the giant star such that the envelope itself overflows its Roche lobe and the companion star accretes mass and launches jets, but the companion is not deep inside the giant envelope, it actually grazes the giant star and the system experiences the grazing envelope evolution (e.g., \citealt{Shiberetal2017, LopezCamaraetal2019, Shiber2018, ShiberSoker2018}). 

We also take the mass loss rate of the giant star to be very high, as expected on the upper asymptotic giant branch (AGB), such that the density in the zone above the giant star is very high. Radiation pressure on dust accelerates the wind. However, the acceleration is less efficient in the case of dense parcels of gas, and they might fall back on to the envelope of the giant star. \cite{Soker2008} termed this extended zone above the surface of giant stars where in addition to the escaping wind there are parcels of gas that do not reach the escape velocity, the \textit{effervescent zone.}
\cite{Soker2008} further argued that accretion from the effervescent zone is efficient and the gas flows through an accretion disk that launches jets. The jets that the more compact companion, basically a main sequence star, launches shape the outflow and hence the descendant planetary nebula. In the present study we simulate the interaction of the jets with the dense wind zone, i.e., the effervescent zone.
 
The idea is that the mass accretion rate is higher than that from a wind because the wind acceleration zone overflows its Roche lobe. \cite{Harpazetal1997} discussed this process. In their words they discussed cases where ``if the companion star is sufficiently close that the Roche lobe of the AGB star moves inside the extended atmosphere, then the slowly moving material will be forced to flow approximately along the critical potential surface (i.e., the Roche lobe) until it flows into the potential lobe of the companion star.'' \cite{MohamedPodsiadlowski2007} and \cite{PodsiadlowskiMohamed2007} (and later papers, e.g., \citealt{MohamedPodsiadlowski2012, Chenetal2017, Saladinoetal2018, Chenetal2020}) studied this process in more details, including 3D hydrodynamical simulations, and termed it a wind-Roche lobe over flow. 
{{{{ \cite{Chenetal2017} find that the mass transfer process in this case is likely to form a circumbinary disk (see also \citealt{Chenetal2020}) as well as an accretion disk around the secondary (see also \citealt{Saladinoetal2018}). }}}}

\cite{AkashiSoker2013} simulated the case where a companion launches jets very close to the surface of an AGB star when the AGB star loses a mass of $\approx 0.1 M_\odot$ in a pulse. The flow is optically thick. We differ by having an optically thin flow and by having a continuous wind. Our simulations are closer to those of \cite{GarciaArredondoFrank2004}.
There are some quantitative differences between the initial conditions in their simulations and ours, as well as two qualitative differences. The first qualitative difference is that we consider radiative cooling while they took a constant adiabatic index of $\gamma=1.01$ to mimic radiative cooling. The second qualitative difference is that they were interested mainly in the shaping of the outflow while we are aiming at revealing the manner by which the companion might enhance mass loss rate from the giant star. For that reason we inject the wind at a velocity smaller than the escape velocity from the AGB star, while they injected the wind at a super-escape velocity. 

We describe the numerical setting in section \ref{sec:numerical}, and our results in section \ref{sec:results}.
In section \ref{sec:summary} we list the main results of our simulations, and further emphasise the important role of jets in the evolution of close binary systems.  

\section{Numerical set up}
\label{sec:numerical}

In this section we describe the features of the simulations that we design to demonstrate the effect of jets on the extended acceleration zone of an intensive wind of an AGB star. 
We run the three-dimensional (3D) hydrodynamical code {\sc pluto} \citep{Mignone2007}.
At the center of the simulation zone $(x,y,z)=(0,0,0)$ we place a spherical AGB star, the primary star, with a radius of $R_1=200\,R_{\sun}$, and mass of $M_1=2\,M_{\sun}$. The full 3D Cartesian grid is taken as a cube with sides  of $4000\,R_{\sun}$ (10 times the diameter of the star). The boundary conditions are transmission for all boundaries. 

In three simulations we employ an adaptive mesh refinement (AMR) grid with three refinement levels, the fiducial value. The base grid resolution is $1/48$ of the grid length (i.e., $83.33 R_{\sun}$), and the finest resolution is $2^3$ times smaller (i.e., $10.4 R_{\sun}$).  
The refinement criterion is the default AMR criterion in {\sc pluto} v. 4.2, 
based on the second derivative error norm \citep{Lohner1987} of the total energy density, 
which effectively tracks the secondary star and the perturbed regions.
In one simulation, the high resolution simulation, we increase the resolution by using a base grid resolution of $1/64$ instead of $1/48$ of the grid length. 
  
Since we do not include the acceleration of the wind, we mimic the acceleration zone of the wind, the effervescent zone, in the following way. The escape velocity from the central star is $v_{\rm esc} = 61.8 \km \s^{-1}$. We inject a spherically symmetric outflow, the wind, at the surface of the AGB star with an initial velocity of $v_{w0} = 60.5 \km \s^{-1}$. We take this outflow velocity to ensure that the wind remains gravitationally bounded to the AGB star, and yet it will exit the grid to avoid a counter inward flow. 
We take the mass loss rate of the primary star to the wind to be $\dot{M_w} = 5 \times 10^{-6}\,M_{\sun} \yr^{-1}$. 
The radial density profile of the wind is given by mass conservation 
\begin{equation}
\rho_w(r) = \frac{\dot{M_w}}{ 4\pi r^2  v_w(r)}, 
\label{eq:density}
\end{equation}
where the velocity of the wind in our simulation (that does not include radiation pressure) is calculated from energy conservation 
\begin{equation}
-\frac{G M_1}{R_1}+\frac{v^2_{w0}}{2} = -\frac{G M_1}{r} + \frac{v^2_w(r)}{2}. 
\label{eq:velocity}
\end{equation}
As the most relevant parts of the wind to the dynamics are the hot parts that have high pressure, we assume for the entire wind an equation of state of a solar composition ionised ideal gas with a mean molecular weight of $\mu =0.62$.  

We include radiative cooling of optically thin gas as implemented in the {\sc pluto} code using the tabulated cooling function from Table 6 in \cite{SutherlandDopita1993}. We do not include dust grains, nor radiation pressure, nor self-gravity of the wind, and nor the gravity of the secondary star that launches the jets. We include only the spherically symmetric gravity of the primary AGB star. 

We assume that a secondary star orbits the primary star with an orbital separation of $a=1000 R_\odot$ and with an orbital period of $T_{\rm orb}=7.1 \yr$, as if the secondary mass is zero. In reality, the secondary star in the scenario we propose has a mass of $M_2 \simeq 0.3-1 M_\odot$. However, since we do not include the gravity of the secondary star, we set its velocity to be as if its mass is zero. 
At that orbit the wind velocity and density (by equations \ref{eq:density} and \ref{eq:velocity}) are $v_w(a) = 24.7 \km \s^{-1}$ and $\rho_w(a) =  2.1 \times 10^{-15} \g \cm^{-3}$. The orbital velocity at that radius is $v_2(a)=19.5 \km \s^{-1}$, and so the relative velocity of the wind and the secondary star is $v_{\rm rel}(a)=31.5 \km \s^{-1}$.

From the above parameters we calculate the Bondi-Hoyle-Lyttleton accretion radius 
to be $R_{\rm acc} = 115 (M_2/0.3 M_\odot) R_\odot$ 
and the accretion rate to be 
\begin{equation}
\dot M_{\rm acc,2}= 2.1 \times 10^{-8}  \left( \frac{M_2}{0.3M_\odot} \right)^2  M_\odot \yr^{-1} .
\label{eq:accretion}
\end{equation}
If a fraction $\eta_j$ of the accreted mas is launched in the jets at about the escape velocity from the secondary star, $v_{\rm jet}=600 \km \s^{-1}$, the power of the jets for our simulated parameters is 
\begin{equation}
\dot{E}_{\rm jets} = 2.4 \times 10^{32}  
\left( \frac{\eta_j}{0.1} \right) 
\left( \frac{M_2}{0.3M_\odot} \right)^2 \erg \s^{-1} .
\label{eq:Ejets}
\end{equation}
 
We assume a lower efficiency than that we derive in equation (\ref{eq:Ejets}), and in the fiducial run we set the mass loss rate into the two jets and the power of the two jets to be $\dot{M}_{\rm jets} = 2.8 \times 10^{-10} M_\odot \yr^{-1}$ 
and $\dot{E}_{\rm jets} = 3.2 \times 10^{31} \erg \s^{-1}$, respectively.   
Namely, we take $\eta_j=0.013$ for $M_2=0.3 M_\odot$. 
We inject the jets in two opposite cones, each extending from the location of the secondary star to a distance of $50 R_\odot$ from it. In the fiducial cases each jet with a half opening angle of $\alpha_j=30^\circ$, 
and with an initial velocity of $v_{\rm j}=600 \km \s^{-1}$.

We run three other cases where in each case we vary one parameter relative to the fiducial run. 
In one case we set $\alpha_j=45^\circ$ instead of $\alpha_j=30^\circ$.
In another case we set the mass loss rate into the jets, hence the energy they carry, to be five times as high, namely, $\dot{E}_{\rm jets,5} = 1.6 \times 10^{32} \erg \s^{-1}$ (still for $\alpha_j=30^\circ$). 
In yet another case we use the same physical parameters as that in the fiducial run, but take a higher resolution in the numerical grid, namely, the base grid resolution is $1/64$ of the grid length instead of $1/48$. 
 
In our flow the radiative cooling is very rapid, and it severely limits the numerical time step. To facilitate a reasonable simulation time, we reduce the cooling rate by a factor of $\eta_{\rm rad} = 1000$. The cooling time in the regions where it is most important is much shorter than the orbital period and/or the adiabatic cooling time. Therefore, this numerical reduction of the cooling rate has a small influence on our main results.  Even with this numerical setup the simulations on our computer cluster lasts for weeks, and so in the present study we limit ourselves to present three cases that demonstrate our idea of jet-driven mass removal in the acceleration zone (effervescent zone) of the wind.  
  
\section{Results}
\label{sec:results}
 
\subsection{Reaching a steady state}
\label{subsec:steady}
\begin{figure} 
\centering
\includegraphics[width=0.45\textwidth]{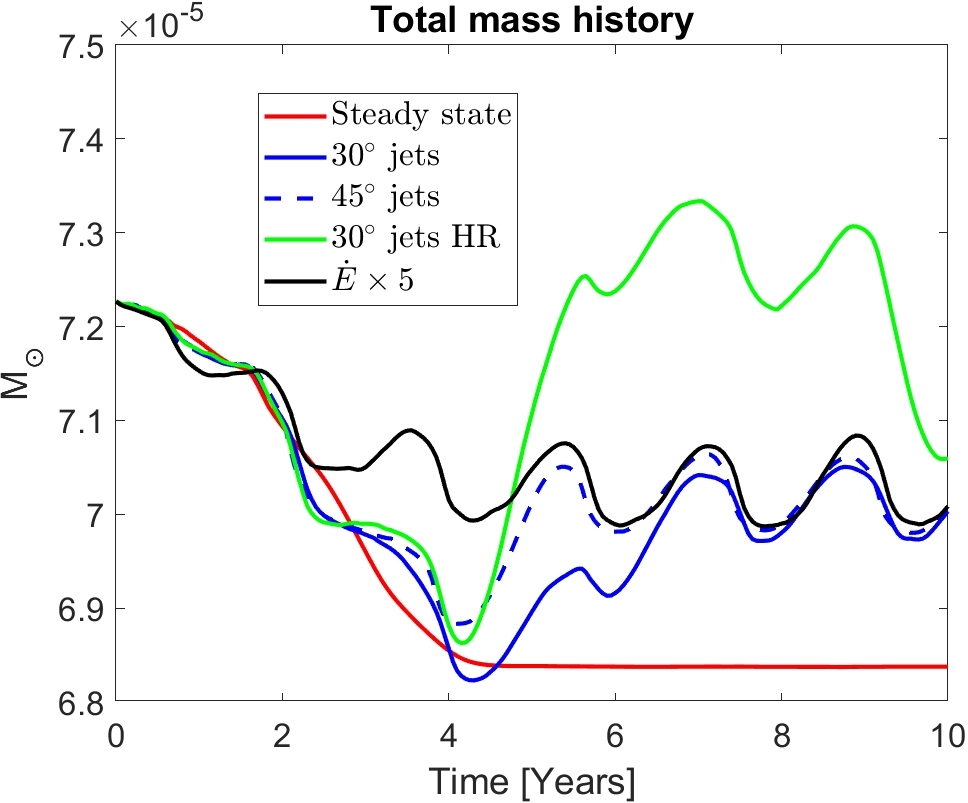}
\caption{The mass inside the grid in the volume $r>300 R_\odot$ as function of time for four simulations with jets, and one without jets. The red line presents the mass for the case without jets to examine that the wind is stable and the flow reaches a steady state. The angles listed in the inset are the half opening angle of the jets, and HR stands for high resolution simulation. The periodic mass variation is due to the cubical boundary of the grid and the spiral structure of the mass that the jets remove. 
}  
\label{fig:MassGrid}
\end{figure}

We first check the time for the flow without jets, namely, only the slow spherical wind, to reach steady state. We inject the spherical wind at $r=300 R_\odot$ and present the mass in the grid above that radius as a function of time by the red line in Fig. \ref{fig:MassGrid}. At early time the initial mass in the grid flows out from the grid at a higher rate than the wind inflow rate. After about 4 years the flow reaches a steady state (line becomes horizontal). We then examine the mass in the grid when we inject jets. The default case is a jets' half opening angle of $\alpha_j=30^\circ$ and our fiducial resolution which we present with a solid blue line. We present also (green line) the mass in the grid for the high resolution simulation (which took a long computational time). 
The periodic variation of the mass in the grid results from the cubical boundary of the grid and the spiral structure of the mass that the jets remove. The dashed-blue line present the mass in the grid for the fiducial resolution but $\alpha_j=45^\circ$, and the black line represents the case with higher jets' energy. The different cases approach a similar behaviour.  

\begin{figure} 
\centering
\includegraphics[width=0.45\textwidth]{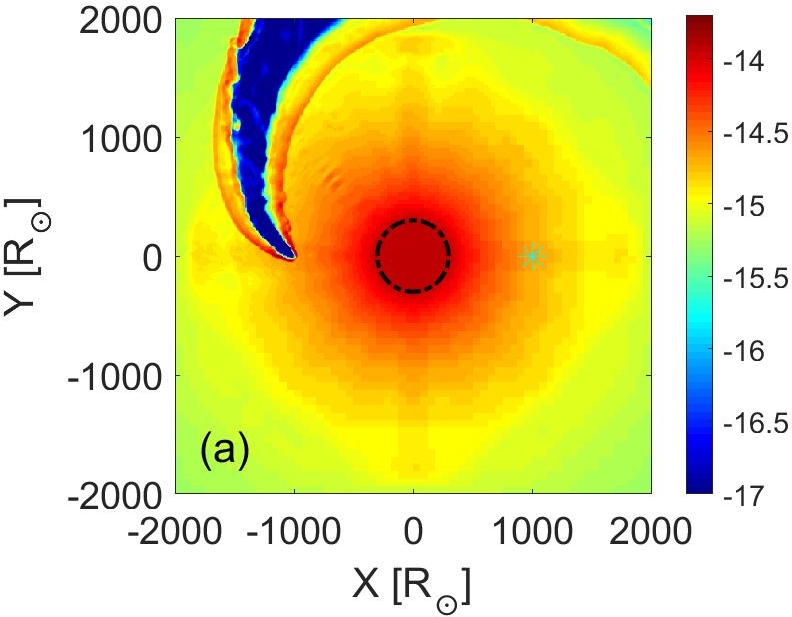}
\includegraphics[width=0.45\textwidth]{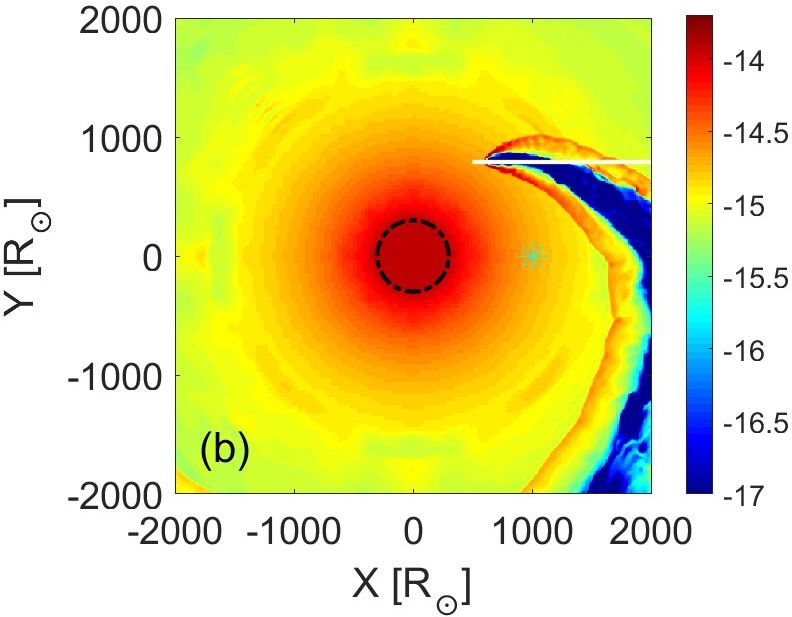}
\includegraphics[width=0.45\textwidth]{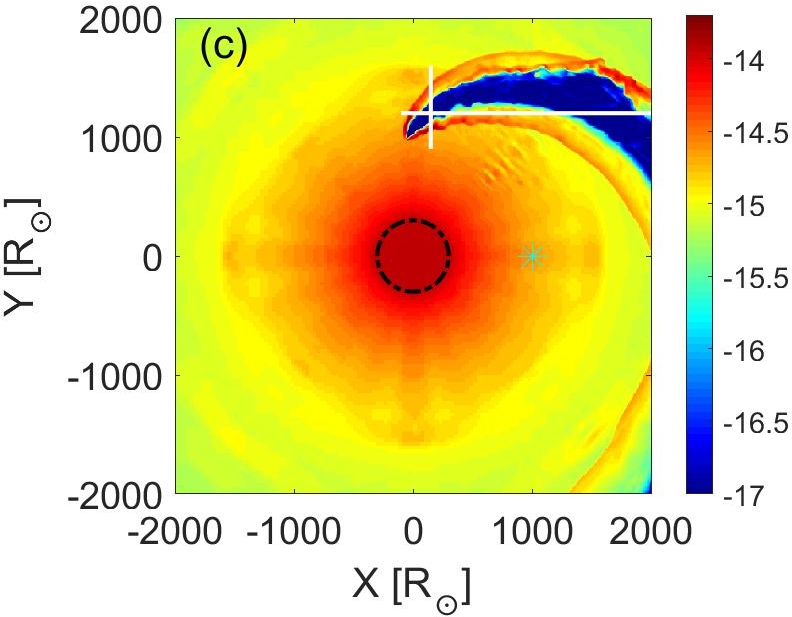}
\caption{Density maps in the equatorial plane ($z=0$) at (a) $t=3.55 \yr$, (b) $t=8.1 \yr$, and (c) $t=8.9 \yr$,
taken from the high-resolution run. 
The colours depict the density in units of $\g \cm^{-3}$ according to the colourbar that is in a logarithmic scale running from $10^{-17} \g \cm^{-3}$ (blue) to $2 \times 10^{-14} \g \cm^{-3}$ (red).
The cyan dot at $(x,y)=(1000R_\odot, 0)$ is the initial location of the secondary star that launches the jets, and the black dashed circle is the spherical surface from where we inject the radial slow wind. Axes are in $R_{\sun}$ units. 
The spiral structure reaches a more or less steady structure at $t \simeq 3 \yr$. 
Cuts along the white lines are presented in Figs. \ref{fig:DensityCuts81} and \ref{fig:DensityCuts89}. }  
\label{fig:Density1}
\end{figure}

In Fig. \ref{fig:Density1} we present the density of the high resolution simulation in the equatorial plane at three times as indicated. Most prominent is the spiral structure that the jets carve in the wind. After about three years the spiral structure reaches a more or less steady state.  We turn to analyse the flow structure on small scales (section \ref{subsec:Instabilities}) and on a large scale (section \ref{subsec:Flow}). 

\begin{figure} 
\centering
\includegraphics[width=0.23\textwidth]{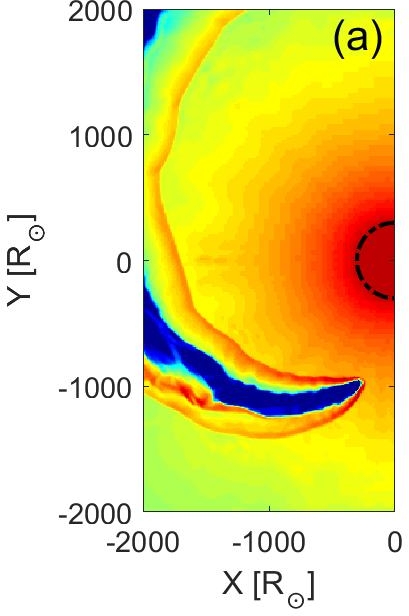}
\includegraphics[width=0.23\textwidth]{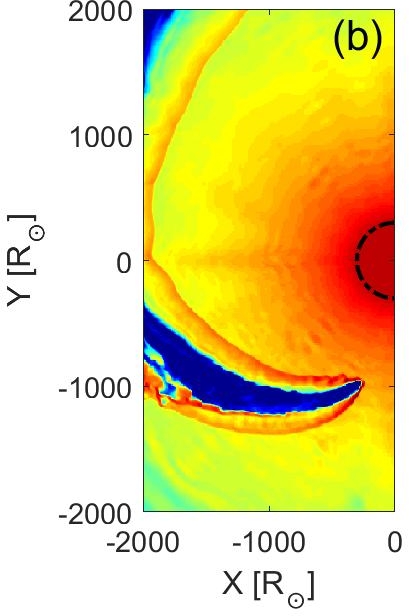}
\caption{A comparison of the density maps at $t=5 \yr$ in the equatorial plane of the
(a) fiducial run, and of the (b) high-resolution
run. The differences are small, thus showing numerical convergence. The density colour coding is according to the colourbar of Fig. \ref{fig:Density1}. 
} 
\label{fig:HR_vs_fiducial}
\end{figure}

In Fig. \ref{fig:HR_vs_fiducial} we compare the fiducial run (left panel a) with the high resolution one (right panel b). 
As expected, the high resolution run better resolves the small scale instabilities and other flow structures of the jets interaction with the dense wind. Beside that, the two simulations are very similar, showing numerical convergence. In the figures we present in this section the structure of the numerical grid is imprinted in the density maps. But the similarity of the results of the two simulations with different resolution show that this has only a small influence on the flow we are focusing on, namely, the jets interaction with the dense wind. 

{{{{ There are some effects that we do not include here, like radiation pressure, the formation of dust, the self gravity of the gas, and the gravity of the secondary star. One of the effects of the secondary stellar gravity is to focus an equatorial outflow, many times in a spiral structure as numerical simulations show (e.g., \citealt{MastrodemosMorris1999, Liuetal2017, MacLeodetal2018, Kimetal2019, Saladinoetal2019, ElMellahetal2020} ). These numerical simulations do not include jets. \cite{Shiberetal2019} include the gravity of the secondary star, the self gravity of the gas, and jets. They obtain a spiral pattern in the equatorial plane as well. However, as they start with the secondary star very close to the surface, in a short time the system enters a common envelope and so the spiral pattern has no time to develop. There is a need for dedicated studies that include both the gravity of the secondary star and  jets.   
  }}}}

\subsection{Instabilities and mixing}
\label{subsec:Instabilities}

\begin{figure} 
\centering
\includegraphics[width=0.45\textwidth]{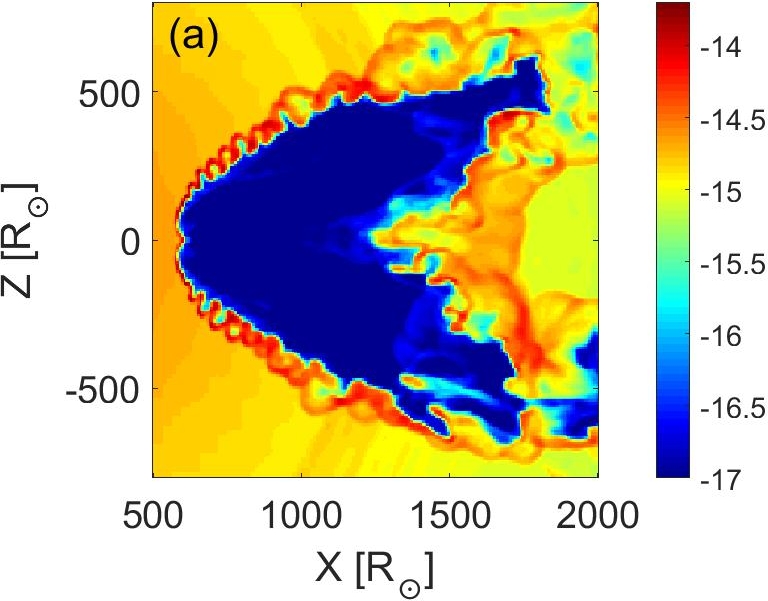}
\includegraphics[width=0.45\textwidth]{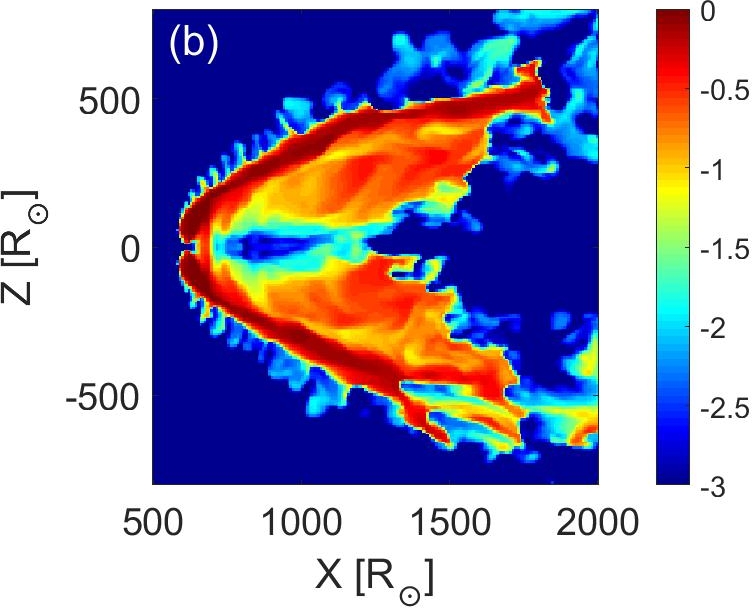}
\caption{Density (upper panel; colour coding according to the colourbar that is in a logarithmic scale running from $10^{-17} \g \cm^{-3}$ (blue) to $2 \times 10^{-14} \g \cm^{-3}$ (red), as in Fig. \ref{fig:Density1}) and jet-tracer (lower panel; colour coding is logarithmic) maps at $t=8.1 \yr$ in the $y=790 R_\odot$ plane that is perpendicular to the equatorial plane as marked by the white line on 
Fig. \ref{fig:Density1}b. 
} 
\label{fig:DensityCuts81}
\end{figure}

\begin{figure} 
\centering
\includegraphics[width=0.24\textwidth]{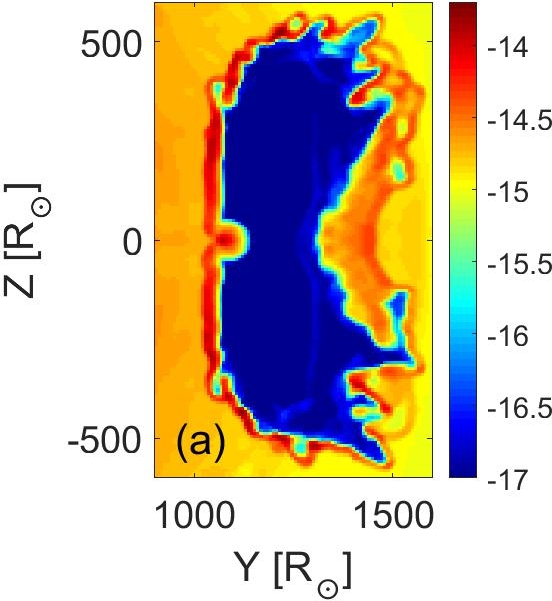}
\includegraphics[width=0.23\textwidth]{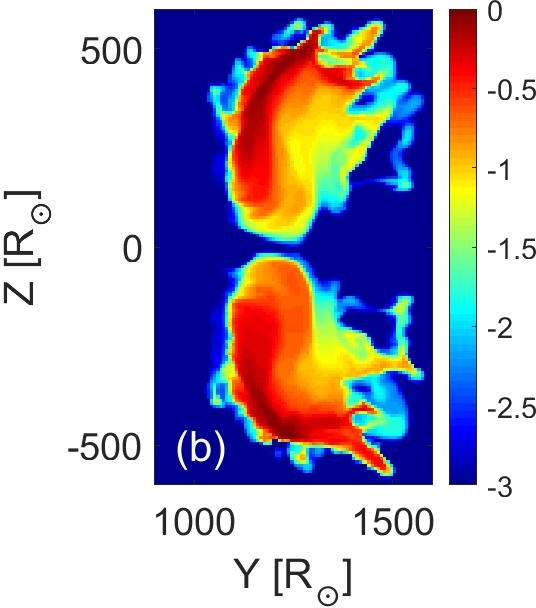}
\includegraphics[width=0.42\textwidth]{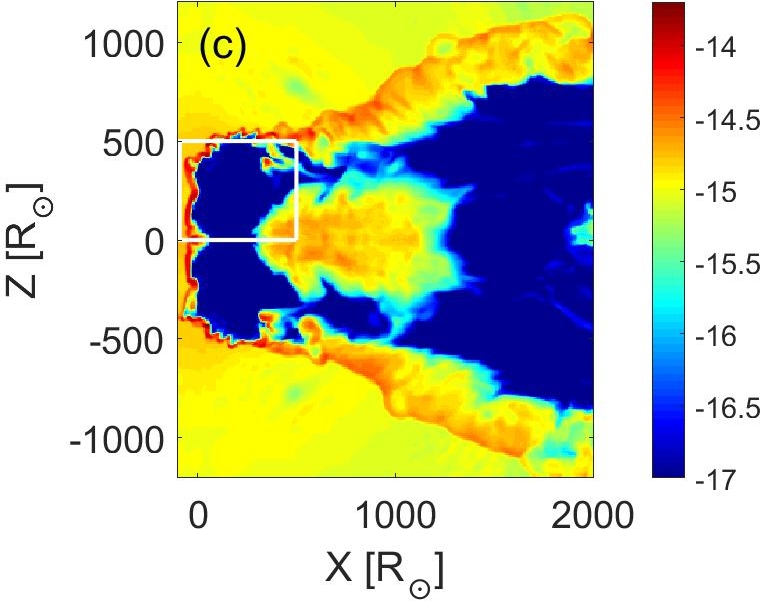}
\includegraphics[width=0.38\textwidth]{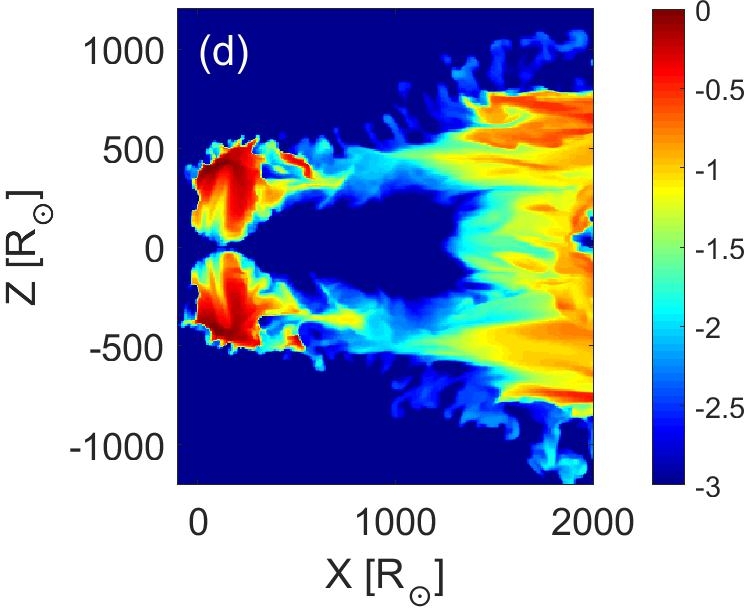}
\caption{Density and jet-tracer maps at $t=8.9 \yr$ in planes perpendicular to the equatorial plane as marked by the white lines on Fig. \ref{fig:Density1}c.
Upper row shows the density (left) and jet-tracer (right) 
in the plane $x=150 R_\odot$, as marked by the vertical solid-white line on Fig. \ref{fig:Density1}c. 
The middle panel shows the density and the lower panel shows the jet-tracer in the plane $y=1200 R_\odot$ as marked by the horizontal white lines on Fig. \ref{fig:Density1}c.
The colourbars for the density and for the tracer scales are as in Fig. \ref{fig:DensityCuts81}.
The white rectangle on panel c refers to the lower panel of 
Fig. \ref{fig:VelocityVectors}.
} 
\label{fig:DensityCuts89}
\end{figure}

In Fig. \ref{fig:DensityCuts81} we present the density and 
jet-tracer maps in the $y=790 R_\odot$ plane and at $t=8.1 \yr$ of the high resolution run, as we mark by the white horizontal line in  Fig. \ref{fig:Density1}b (middle panel). 
The jet-tracer variable follows the gas that was injected in the jets, and in each grid cell it is equal to the fraction of the gas that originated in the jets.
In Fig. \ref{fig:DensityCuts89} we present the density and jet-tracer maps in two cuts perpendicular to the orbital plane at $t=8.9 \yr$.  The two panels in the upper row are in the $x=150 R_\odot$ plane that we mark by the white vertical line in Fig. \ref{fig:Density1}c (lower panel), while the two lower panels are for the $y=1200 R_\odot$ plane as marked by the white horizontal line in  Fig. \ref{fig:Density1}c. 

Figs. \ref{fig:DensityCuts81} and \ref{fig:DensityCuts89} emphasise the Kelvin-Helmholtz instabilities that develop in the shear layer between the jets and the envelope. These instabilities mix the gas that originated in the jets with the gas in the wind, as the jet-tracer maps show. The mixing not only transfers gas from the jets to the wind, but also transfers energy from the jets to the wind to unbind some of the gas in the acceleration zone of the wind (section \ref{subsec:Flow}). The mixing is induced by the complicated flow pattern, like vortexes, that results from instabilities. \cite{HillelSoker2014} studied and discussed the efficient transport of energy from jets to the ambient gas that results from mixing by vortexes. Although they discussed jets in cooling flows in clusters of galaxies, the physics is similar in our flow setting.

\begin{figure} 
\centering
\includegraphics[width=0.45\textwidth]{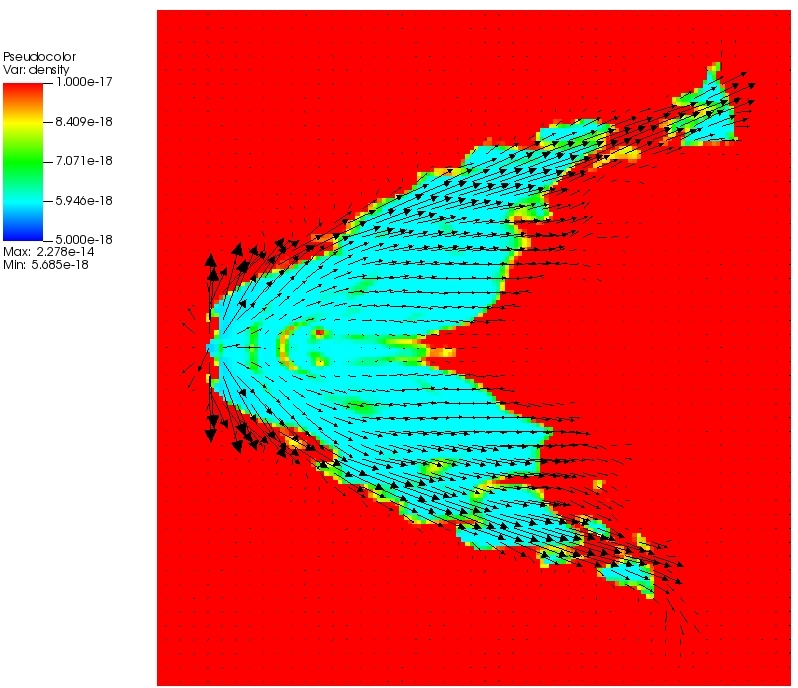}
\includegraphics[width=0.45\textwidth]{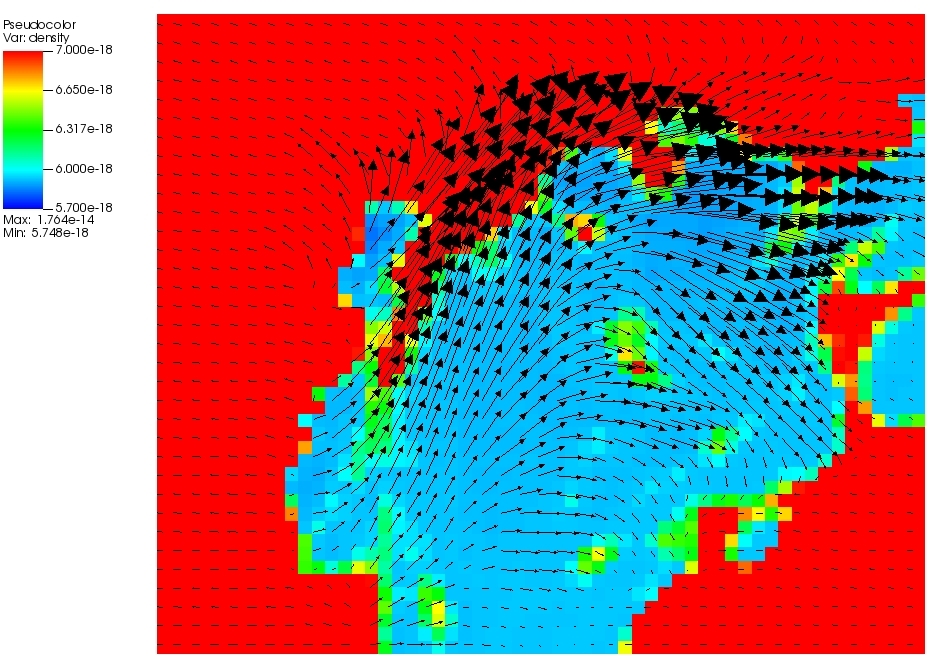}
\caption{Velocity arrows on top of the density maps. The lengths of the arrows proportional to the velocity, with a maximum value of $650 \km \s^{-1}$. Upper panel: The same region and time as in Fig.  \ref{fig:DensityCuts81} ($t=8.1 \yr$; $y=790 R_\odot$; $500 R_\odot \le x \le 2000 R_\odot$; $-800 R_\odot \le z \le 800 R_\odot$).
Lower panel: The same plane as in Fig. \ref{fig:DensityCuts89}c, but only the region marked by the white rectangle there ($t=8.9 \yr$; $y=1200 R_\odot$; $-100 R_\odot \le x \le 500 R_\odot$; $0 R_\odot \le z \le 500 R_\odot$).
The arrows demonstrate the complex flow structure of the disturbed wind zone.
} 
\label{fig:VelocityVectors}
\end{figure}

In Fig. \ref{fig:VelocityVectors} we present velocity maps. The upper panel presents the flow in the same region as in fig. \ref{fig:DensityCuts81} and at $t=8.1 \yr$. The lower panel presents the velocity map of the white rectangle marked on Fig. \ref{fig:DensityCuts89}c. These panels demonstrate the complicated flow pattern that results from the instabilities. 


Figure \ref{fig:DensityCuts81}-\ref{fig:VelocityVectors} demonstrate the complex flow interaction between the jets and the wind, and the development of instabilities and vortexes. 
From Figs. \ref{fig:DensityCuts81} and \ref{fig:DensityCuts89} we find that the typical distance between `instability fingers' is $\approx 50 R_\odot$. This is about the size of the region from where we inject the jets (the length of each of the two opposite cones). 
Our experience with the hydrodynamical code {\sc pluto} \citep{RefaelovichSoker2012} shows that the size of the dominant  Kelvin-Helmholtz modes, e.g., size of vortexes (or turbulent eddies), are somewhat larger than the cross section of the jets. 

\subsection{Flow properties}
\label{subsec:Flow}
 
\begin{figure} 
\centering
\includegraphics[width=0.45\textwidth]{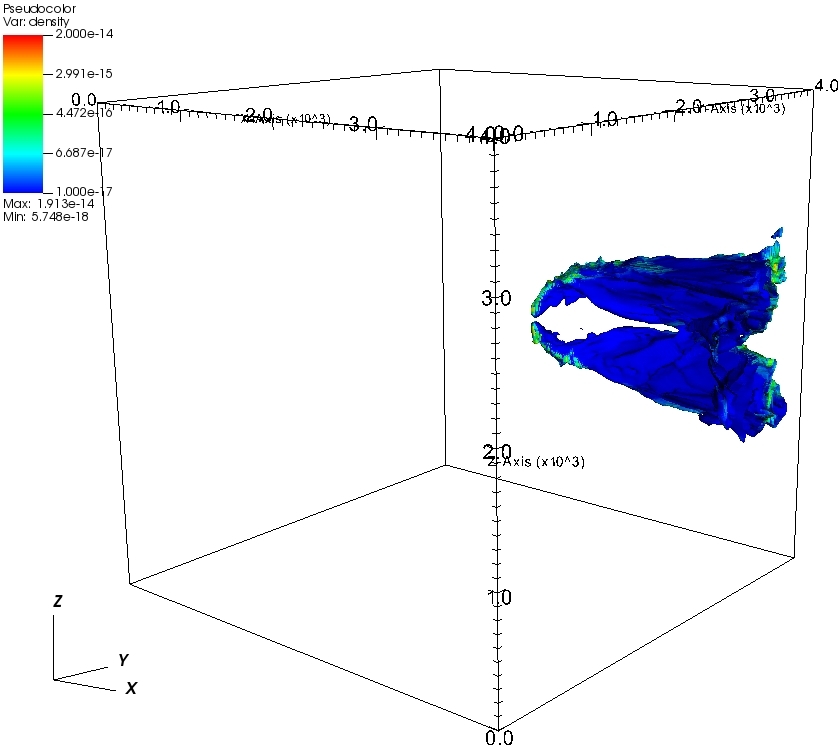}
\includegraphics[width=0.45\textwidth]{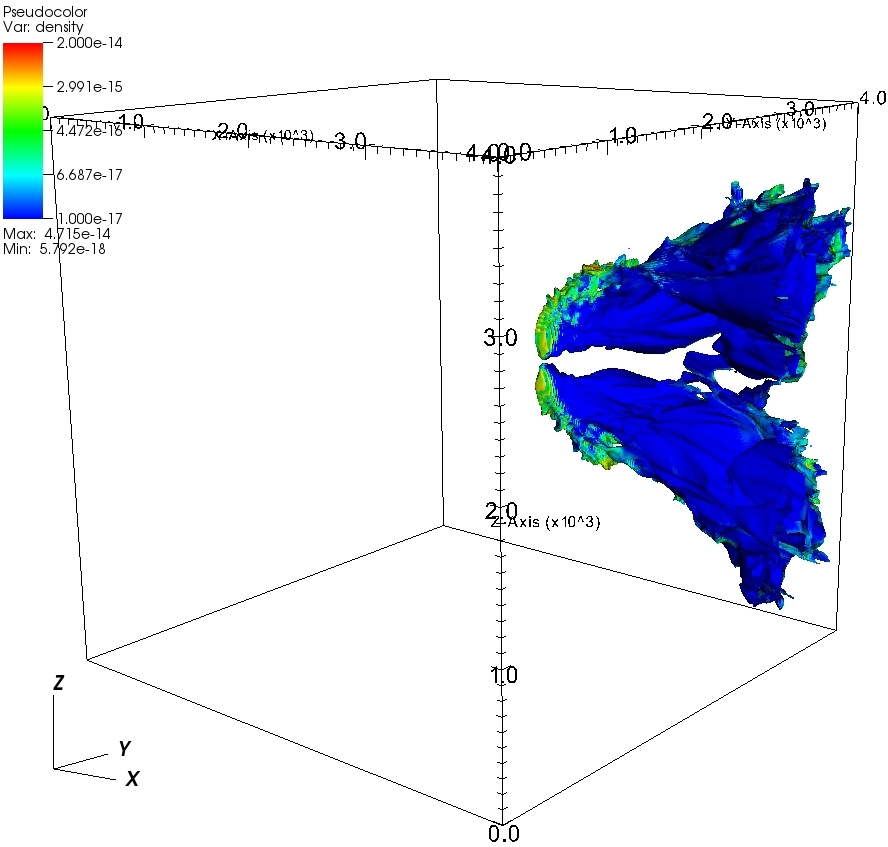}
\caption{Density-surface 3D maps of the gas that originated in the jets and mixed with the wind so the jet-tracer value is $>10^{-4}$. 
Colours are density surfaces according to the colour bars on the left from $10^{-17} \g \cm^{-1}$ (blue) to $2 \times 10^{-14} \g \cm^{-1}$ (red). 
Top panel shows the fiducial case with $\alpha_j=30^\circ$ and the bottom panel shows the case with 5 times as much jets' energy. 
Both maps were taken at $t=8.9 {~\rm years}$
}  
\label{fig:3D_tracer2}
\end{figure}

We turn to discuss the global properties of the flow. In Fig. \ref{fig:3D_tracer2} we present density-surface 3D maps of the gas that originated in the jets and mixed with the wind. The figure show surfaces of constant densities, but only for regions where the jet-tracer is $>10^{-4}$; namely, only computational zones into which some jets' gas is mixed.
The upper panel presents the fiducial run,
whereas the lower panel presents high-energy run where the mass loss rate into the jets is five times larger. As expected, the jets inflate larger bubbles. 
This figure demonstrates the bending of the jets and the interaction region where the jets mix with the wind gas. The jets add energy to the wind and unbind part of the wind (boost its energy to be positive) that leaves the grid in a spiral pattern (Fig. \ref{fig:Density1}). 

\begin{figure} 
\centering
\includegraphics[width=0.45\textwidth]{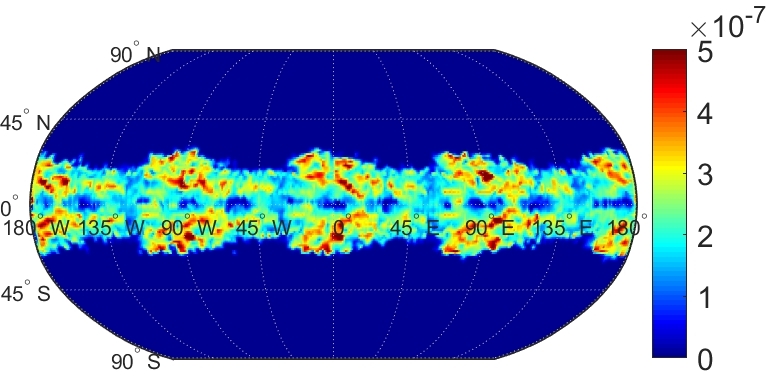}
\includegraphics[width=0.45\textwidth]{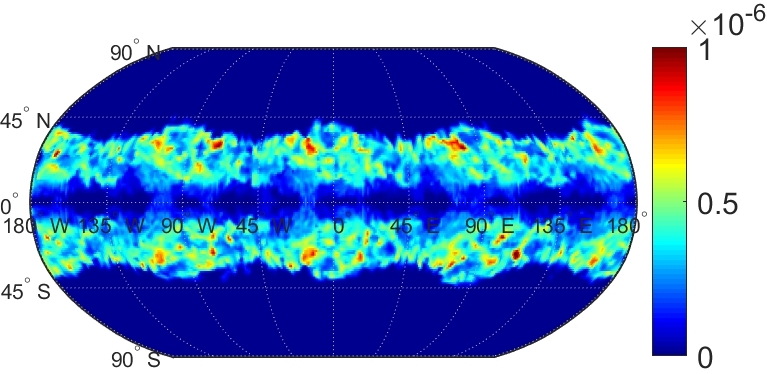}
\caption{Globe-maps of the unbound average mass flux 
through a sphere of radius $r=1900 R_{\sun}$
at $t=8.9 {~\rm years}$ for the fiducial case with $\alpha_j = 30^\circ$ and for the
the case with 5 times as much jets' energy.
The colour coding is in units of $M_\odot \yr^{-1}$, differ for the two panels,
and the value is as if the entire sphere (solid angle of $4 \pi$) would have the same mass flux as that through the given direction.
} 
\label{fig:Globe-map}
\end{figure}

In Fig. \ref{fig:Globe-map} we present two Globe-maps of the average outflow flux of the unbound gas as function of direction and through a sphere of radius $R_{\rm out}=1900 R_\odot$. 
We average over one orbital period from $t=2.9 \yr$ to $t=10 \yr$, for the fiducial run with $\alpha_j=30^\circ$ (upper map), and
 for the 
higher-energy case (lower map).
The colour coding of the two panels is different and in units of $M_\odot \yr^{-1}$, 
and the value is as if the entire sphere (solid angle of $4 \pi$) would have the same mass flux as that through the given direction.
This figure quantitatively presents what earlier figures show qualitatively. 
Namely, that the unbound material leaves the grid in a flow around the equatorial plane.
As the jets' energy increases, the two jets
are seperated. 
In Fig. \ref{fig:Globe-map} we see that the outflow pattern has a periodic variation with $\theta$ (longitude) every 90 degrees. The reason is the Cartesian structure of cells in the numerical grid. As we discussed earlier, there are many instability modes that develop to vortexes (turbulent eddies) and further cause the wiggling of the spiral structure. The initial perturbations that develop to these non-linear structures are caused by the finite-resolution grid, and hence the final non-linear structure has the imprint of the Cartesian grid. 
   
\begin{figure} 
\centering
\includegraphics[width=0.45\textwidth]{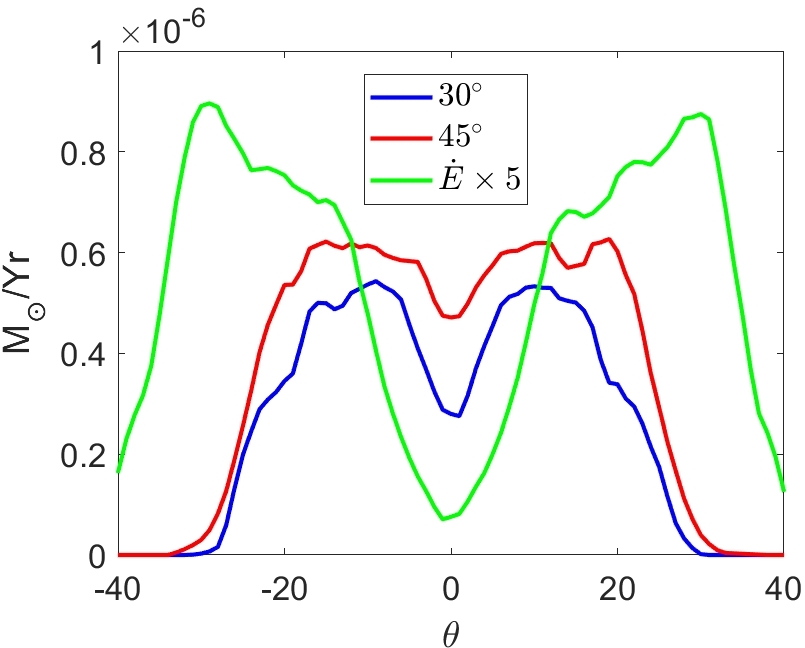}
\caption{The $\phi$-averaged unbound outflow mass flux. 
We average over the angle $\phi$ from $\phi=0^\circ$ to $\phi=360^\circ$ to obtain the dependence of the unbound outflow flux on latitude. The units in $M_\odot \yr^{-1}$ are as if the entire sphere (solid angle of $4 \pi$) would have the same unbound outflow mass flux as that of the average at the given angle $\theta$. 
The blue line is for the fiducial case with $\alpha_j=30^\circ$, the red line is for the $\alpha_j=45^\circ$ case, and the green line is for $\alpha_j=30^\circ$ and 5 times energy in the jets. } 
\label{fig:IntPhi}
\end{figure}

In Figure \ref{fig:IntPhi} we average unbound mass outflow flux over the angle $\phi$ from $\phi=0^\circ$ to $\phi=360^\circ$ to obtain the dependence of the unbound outflow flux on latitude. The units are as if the entire sphere (solid angle of $4 \pi$) would have the same unbound outflow mass flux as that of the average at the given angle $\theta$. We present the graphs for the fiducial case with $\alpha_j=30^\circ$, for the $\alpha_j=45^\circ$ case, and for $\alpha_j=30^\circ$ but 5 times as much energy. Fig. \ref{fig:IntPhi} shows that the jets eject mass mainly around the equatorial plane, with a lower value very close to the equatorial plane. The maximum outflow flux of unbound material is at an angle of $\simeq 10^\circ$ from the equatorial plane for the two cases with the fiducial energy, and at higher latitudes for the higher energy simulation, as we see also in Fig. \ref{fig:Globe-map}.      

\begin{figure} 
\centering
\includegraphics[width=0.40\textwidth]{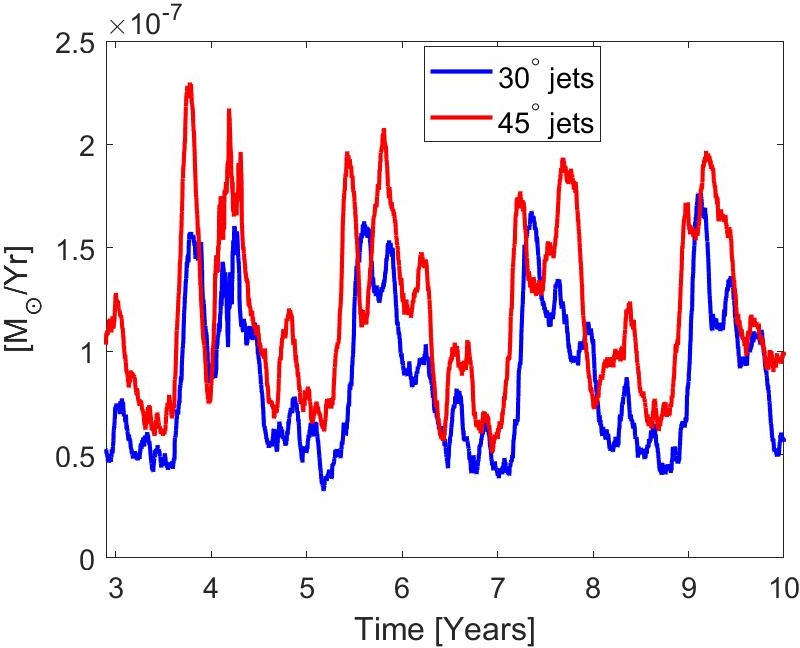}
\caption{The unbound mass outflow rate through 
a sphere of $r=1900 R_{\sun}$ 
as function of time, for two cases as indicated.
}  
\label{fig:mass-flux}
\end{figure}

Figures \ref{fig:Density1}-\ref{fig:Globe-map} show also that the jet-wind interaction spiral region is very clumpy. In addition, the spiral pattern does not have a constant shape, but it rather wiggles slightly. Both these effects cause the mass flux through a given sphere to fluctuate. We demonstrate this in Fig. \ref{fig:mass-flux} that shows the unbound mass outflow rate through a sphere of radius $R_{\rm out} = 1900 R_\odot$ as function of time. There are large fluctuations of the mass flux. A typical time span between peaks of fluctuations is $\Delta t _{\rm fluc}  \approx 1 \yr$. This time scale can be understood as the width of the spiral pattern at a radius of $r=1900 R_\odot$, $\Delta r \simeq 800 R_\odot$ divided by the radial velocity of the spiral pattern, $v_{\rm s,r} \simeq 18 \km \s^{-1}$. 

We recall that in our simulations the wind by itself has a negative energy, namely, it is bound, but only marginally so. Fig. \ref{fig:mass-flux} shows the mass loss rate of the unbound wind material to be $\simeq 10^{-7} M_\odot \yr^{-1}$. The jets accelerate this mass to escape velocities. The mass loss rate into the jets in the simulations we present in Fig. \ref{fig:mass-flux} is $\dot{M}_{\rm jets} = 2.8 \times 10^{-10} M_\odot \yr^{-1}$, less than a percent of the unbound wind mass loss rate. This shows the jets to be efficient in unbinding gas from the acceleration zone of the wind.   

We end the discussion of the flow by referring to our reduction of the cooling rate by a factor of $\eta_{\rm rad} = 1000$ due to severe numerical limitations on the time step (section \ref{sec:numerical}). 
We find the post shock temperature of the jets to be $T_{\rm j,s} \simeq 3 \times 10^6 \K$. Because of the oblique shock, in most of the jets' post-shock volume the velocity is faster than the post-shock velocity of a perpendicular shock ($v_{\rm j}/4$), and its value is $v_{\rm j,s} \simeq 500 \km \s^{-1}$. The flow time scale over a typical region of $D=500 R_\odot$ is then $\tau_{j,f} \simeq D/v_{\rm j,s} \approx 0.02 \yr$. The jets' post-shock density is $\rho_{\rm j,s} \simeq 10^{-17} \g \cm^{-1}$, from which we calculate the radiative cooling time from a temperature of $T_{\rm j,s} \simeq 3 \times 10^6 \K$ to be $\tau_{\rm j,rad} \approx 0.1 \yr$. Since we decrease the radiative cooling time by a large factor, radiative cooling of this gas is negligible in our numerical setting. However, adiabatic cooling is more important than radiative cooling at these temperatures, so that the effect of reducing cooling time is small. At lower temperatures of $\approx 10^5 \K$ the adiabatic cooling is longer due to slower flows. However, at these low temperatures and higher densities the radiative cooling time even with our reduction of radiative cooling rate by a factor of $\eta_{\rm rad} = 1000$ is still short enough to play a role, i.e., much shorter than the flow time. 
To summarise, although not completely accurate, our numerical scheme of reduced radiative cooling rate still captures the general flow properties.   
  
\section{DISCUSSION AND SUMMARY}
\label{sec:summary}
 
The main new result of our study is the introduction of the notion that jets can remove bound mass from the extended acceleration zone of the wind of AGB stars that experience a high mass loss rate. We did not include any calculation of the acceleration of the wind to overcome the gravity of the AGB star, nor did we calculate the formation of an extended zone above the AGB star where a large fraction of the mass is still bound, and therefore  falls back (the \textit{effervescence zone}). Therefore, we could not accurately quantify the effect of mass removal from the effervescence zone by jets. We could only present crude estimates, that nonetheless, show that this process might be significant for the evolution of some binary AGB stars.   

The second finding, similar to, e.g., the grazing envelope evolution simulations of \cite{ShiberSoker2018}, is the bending of the jets and the wind material it mixes with toward the equatorial plane (Fig. \ref{fig:3D_tracer2}). This leads to a mass loss that is concentrated near the equatorial plane (Figs. \ref{fig:Globe-map}, \ref{fig:IntPhi}). 

The third finding, similar to our finding of simulations of jets interacting with the envelope of an AGB star \citep{Schreieretal2019}, 
is the development of instabilities, i.e., density fluctuations (Figs. \ref{fig:DensityCuts81} and \ref{fig:DensityCuts89}), and turbulent flow (Fig. \ref{fig:VelocityVectors}), in the interaction zones of the jets with the wind. This turbulence leads to mixing of the jets with the wind, and therefore to the transfer of energy from the jets to the wind, a process that unbinds part of the wind. The spiral pattern of the gas that the jets unbinds (Fig. \ref{fig:Density1}) and the instabilities lead to a fluctuating mass loss rate through a given sphere (Fig. \ref{fig:mass-flux}). 
    
We place our study in a broader contest. 
There is a very rich variety of binary interaction types that lead to the formation of non-spherical nebulae around evolved stars, particularly planetary nebulae, e.g., some very recent simulations of jet shaping \citep{AkashiSoker2018, EstrellaTrujilloetal2019, RechyGarciaetal2019, Balicketal2019, Zouetal2020}. For the many binary (or triple) interaction types (or evolutionary routes), most of them must be rare, and some are even very rare. Accompanying these different types of binary interactions is a rich variety of processes, including tidal interaction, the presence of a tertiary star, mass transfer, winds, and jets that a mass-accreting companion might launch. 
Although each of the (very) rare evolutionary routes accounts for only one to a few observed systems (or even none that has been observe yet), the study of many (very) rare evolutionary routes, some of them with rare processes, will eventually build a full picture of binary interaction of evolved stars. 

In the present study we simulated such a rare evolutionary route where a companion is present in the extended wind acceleration zone of an upper AGB star that experiences a high mass loss rate. Because of the high mass loss rate we assumed (see section \ref{sec:intro}) that there is an extended wind acceleration zone above the AGB star, where many parcels of gas fall back onto the AGB envelope (a zone that \citealt{Soker2008} termed the \textit{effervescent zone}). We further assumed that the companion main sequence star accretes mass and launches jets. In the present study we simulate a few cases of such a possible interaction, that we consider rare. 

Since we assumed a spherically symmetric wind, the jets did not break out along the polar directions. We did not consider the process where the wind acceleration zone overflows its Roche lobe (\citealt{Harpazetal1997, MohamedPodsiadlowski2007, PodsiadlowskiMohamed2007, MohamedPodsiadlowski2012, Chenetal2017, Saladinoetal2018, Chenetal2020}) such that the accretion flow from near the equatorial plane is much larger while the wind along the polar direction is weaker. In such a case the jets are stronger due to higher accretion rate onto the companion, and the wind polar density is lower, and the jets might break from the polar directions and shape a bipolar outflow with two opposite large lobes. This, also rare, type of interaction requires a study of its own.  

Our results have implications also to eccentric binary systems where near periastron passages the jets that the companion launches remove mass from the envelope of the AGB star. Our results show that the jets can remove some mass also near apastron passages, although much less than from the envelope. Future studies should explore such eccentric binary systems while considering our results.

\section*{Acknowledgments}

This research was supported by a generous grant from Prof. Amnon Pazy Research Foundation.
N.S. research is partially supported by the Charles Wolfson Academic Chair.

\label{lastpage}

\begin{thebibliography}{}

\bibitem[Akashi \& Soker(2013)]{AkashiSoker2013} Akashi, M., \& Soker, N.\ 2013, \mnras, 436, 1961 

\bibitem[Akashi, \& Soker(2018)]{AkashiSoker2018} Akashi, M., \& Soker, N.\ 2018, \mnras, 481, 2754

\bibitem[Armitage \& Livio(2000)]{ArmitageLivio2000} Armitage, P.~J., \& Livio, M.\ 2000, \apj, 532, 540

\bibitem[Balick et al.(2019)]{Balicketal2019} Balick, B., Frank, A., \& Liu, B.\ 2019, arXiv e-prints, arXiv:1911.12812

\bibitem[Chamandy et al.(2018)]{Chamandyetal2018} Chamandy, L., Frank, A., Blackman, E.~G., et al.\ 2018, \mnras, 480, 1898

\bibitem[Chen et al.(2017)]{Chenetal2017} {{{{ Chen, Z., Frank, A., Blackman, E.~G., Nordhaus, J., \& Carroll-Nellenback J.\ 2017, \mnras, 468, 4465 }}}}

\bibitem[Chen et al.(2020)]{Chenetal2020} {{{{ Chen, Z., Ivanova, N., \& Carroll-Nellenback, J.\ 2020, arXiv e-prints, arXiv:1910.08027 }}}} 

\bibitem[Chevalier(2012)]{Chevalier2012} Chevalier, R.~A.\ 2012, \apj, 752, L2

\bibitem[Danieli \& Soker(2019)]{DenieliSoker2019} Danieli, B., \& Soker, N.\ 2019, \mnras, 482, 2277

\bibitem[El Mellah et al.(2020)]{ElMellahetal2020} El Mellah, I., Bolte, J., Decin, L., Homan, W., \& Keppens, R., 2020, arXiv, arXiv:2001.04482

\bibitem[Estrella-Trujillo et al.(2019)]{EstrellaTrujilloetal2019} Estrella-Trujillo, D., Hern{\'a}ndez-Mart{\'\i}nez, L., Vel{\'a}zquez, P.~F.,Esquivel, A., \& Raga, A.~C.\ 2019, \apj, 876, 29

\bibitem[Garc{\'{\i}}a-Arredondo \& Frank(2004)]{GarciaArredondoFrank2004} Garc{\'{\i}}a-Arredondo, F., \& Frank, A.\ 2004, \apj, 600, 992 

\bibitem[Gilkis et al.(2019)]{Gilkisetal2019} Gilkis, A., Soker, N., \& Kashi, A.\ 2019, \mnras, 482, 4233

\bibitem[Gorlova et al.(2015)]{Gorlovaetal2015} Gorlova, N., Van Winckel, H., Ikonnikova, N.~P., Burlak, M.~A., Komissarova, G.~V., Jorissen, A., Gielen, C., Debosscher, J., \& Degroote, P. 2015\, \mnras, 451, 2462
 
\bibitem[Harpaz et al.(1997)]{Harpazetal1997} Harpaz, A., Rappaport, S., \& Soker, N.\ 1997, \apj, 487, 809 

\bibitem[Hillel, \& Soker(2014)]{HillelSoker2014} Hillel, S., \& Soker, N.\ 2014, \mnras, 445, 4161

\bibitem[Kim et al.(2019)]{Kimetal2019} Kim, H., Liu, S.-Y., \& Taam, R.~E.\ 2019, \apjs, 243, 35

\bibitem[Liu et al.(2017)]{Liuetal2017} Liu, Z.-W., Stancliffe, R.~J., Abate, C., \& Matrozis E.,.\ 2017, \apj, 846, 117

\bibitem[Lohner(1987)]{Lohner1987} Lohner, R.\ 1987, Computer Methods in Applied Mechanics and Engineering, 61, 323

\bibitem[L{\'o}pez-C{\'a}mara et al.(2019)]{LopezCamaraetal2019} L{\'o}pez-C{\'a}mara, D., De Colle, F., \& Moreno M{\'e}ndez, E.\ 2019, \mnras, 482, 3646

\bibitem[MacLeod et al.(2018)]{MacLeodetal2018} MacLeod, M., Ostriker, E.~C., \& Stone, J.~M.\ 2018, \apj, 868, 136
 
\bibitem[MacLeod \& Ramirez-Ruiz(2015)]{MacLeodRamirezRuiz2015} MacLeod, M., \& Ramirez-Ruiz, E.\ 2015, \apj, 803, 41

\bibitem[Mastrodemos \& Morris(1999)]{MastrodemosMorris1999} Mastrodemos, N., \& Morris, M.\ 1999, \apj, 523, 357

\bibitem[Mignone et al.(2007)]{Mignone2007} Mignone, A., Bodo, G., Massaglia, S., et al.\ 2007, \apjs, 170, 228

\bibitem[Moreno M{\'e}ndez et al.(2017)]{MorenoMendezetal2017} Moreno M{\'e}ndez, E., L{\'o}pez-C{\'a}mara, D., \& De Colle, F.\ 2017, \mnras, 470, 2929 

\bibitem[Mohamed \& Podsiadlowski(2007)]{MohamedPodsiadlowski2007} Mohamed, S., \& Podsiadlowski, P.\ 2007, 15th European Workshop on White Dwarfs, 372, 397 

\bibitem[Mohamed \& Podsiadlowski(2012)]{MohamedPodsiadlowski2012} Mohamed, S., \& Podsiadlowski, P.\ 2012, Baltic Astronomy, 21, 88

\bibitem[Papish et al.(2015)]{Papishetal2015} Papish, O., Soker, N., \& Bukay, I.\ 2015, \mnras, 449, 288

\bibitem[Podsiadlowski \& Mohamed(2007)]{PodsiadlowskiMohamed2007} Podsiadlowski, P., \& Mohamed, S.\ 2007, Baltic Astronomy, 16, 26

\bibitem[Rechy-Garc{\'\i}a et al.(2019)]{RechyGarciaetal2019} Rechy-Garc{\'\i}a, J.~S., Pe{\~n}a, M., \& Vel{\'a}zquez, P.~F.\ 2019, \mnras, 482, 1163
 
\bibitem[Refaelovich, \& Soker(2012)]{RefaelovichSoker2012} Refaelovich, M., \& Soker, N.\ 2012, \apjl, 755, L3

\bibitem[Ricker \& Taam(2012)]{RickerTaam2012} Ricker, P.~M., \& Taam, R.~E.\ 2012, \apj, 746, 74

\bibitem[Sabach et al.(2017)]{Sabachetal2017} Sabach, E., Hillel, S., Schreier, R., \& Soker, N.\ 2017, \mnras, 472, 4361

\bibitem[Saladino et al.(2019)]{Saladinoetal2019} Saladino, M.~I., Pols, O.~R., \& Abate, C.\ 2019, \aap, 626, A68

\bibitem[Saladino et al.(2018)]{Saladinoetal2018} {{{{ Saladino, M.~I., Pols, O.~R., van der Helm, E., Pelupessy, I., \& Portegies Zwart S.\ 2018, \aap, 618, A50 }}}}

\bibitem[Schreier et al.(2019)]{Schreieretal2019} Schreier, R., Hillel, S., \& Soker, N.\ 2019, \mnras, 490, 4748

\bibitem[Shiber(2018)]{Shiber2018} Shiber, S.\ 2018, Galaxies, 6, 96

\bibitem[Shiber et al.(2019)]{Shiberetal2019} Shiber, S., Iaconi, R., De Marco, O., \& Soker, N.\ 2019, \mnras, 1953

\bibitem[Shiber et al.(2017)]{Shiberetal2017} Shiber, S., Kashi, A., \& Soker, N.\ 2017, \mnras, 465, L54

\bibitem[Shiber et al.(2016)]{Shiberetal2016} Shiber, S., Schreier, R., \& Soker, N.\ 2016, Research in Astronomy and Astrophysics, 16, 117

\bibitem[Shiber \& Soker(2018)]{ShiberSoker2018} Shiber, S., \& Soker, N.\ 2018, \mnras, 477, 2584 

\bibitem[Soker(2004)]{Soker2004} Soker, N.\ 2004, \na, 9, 399

\bibitem[Soker(2008)]{Soker2008} Soker, N.\ 2008, \na, 13, 491

\bibitem[Soker(2016a)]{Soker2016Rev} Soker, N.\ 2016a, New Astronomy REviews, 75, 1

\bibitem[Soker \& Gilkis(2018)]{SokerGilkis2018} Soker, N., \& Gilkis, A.\ 2018, \mnras, 475, 1198

\bibitem[Staff et al.(2016a)]{Staffetal2016MN} Staff, J.~E., De Marco, O., Macdonald, D., Galaviz, P., Passy, J.C., Iaconi, R., \& Mac Low, M.-M\ 2016a, \mnras, 455, 3511

\bibitem[Sutherland \& Dopita(1993)]{SutherlandDopita1993} Sutherland, R.~S., \& Dopita, M.~A.\ 1993, \apjs, 88, 253

\bibitem[Thomas et al.(2013)]{Thomasetal2013} Thomas, J.~D., Witt, A.~N., Aufdenberg, J.~P., Bjorkman, J. E., Dahlstrom, J. A., Hobbs, L. M., \& York, D. G.\ 2013, \mnras, 430, 1230

\bibitem[Van Winckel(2017)]{VanWinckel2017} Van Winckel, H.\ 2017, in Planetary Nebulae: Multi-Wavelength Probes of Stellar and Galactic Evolution, Proceedings IAU Symposium No. 323,  p. 231, eds. X. Liu, L. Stanghellini, and A. Karakas A.C.

\bibitem[Witt et al.(2009)]{Wittetal2009} Witt, A.~N., Vijh, U.~P., Hobbs, L.~M., Aufdenberg, J. P., Thorburn, J. A., \& York, D. G.\ 2009, \apj, 693, 1946

\bibitem[Zou et al.(2019)]{Zouetal2020} Zou, Y., Frank, A., Chen, Z., et al.\ 2019, arXiv e-prints, arXiv:1912.01647

\end{thebibliography}
\end{document}